\documentclass[11pt,twoside]{article}
\usepackage{fleqn,espcrc1}
\usepackage{amssymb}
\newcommand{\be}{\begin{equation}}
\newcommand{\ee}{\end{equation}}
\newcommand{\ba}{\begin{eqnarray}}
\newcommand{\ea}{\end{eqnarray}}
\renewcommand{\thefootnote}{\fnsymbol{footnote}}
\title{\bf \LARGE Nonlinear Supersymmetric (Darboux) Covariance of the Ermakov-Milne-Pinney Equation}
\author{M. V. Ioffe\address[io]{Department of Theoretical Physics, Institute of Physics, University
of Sankt-Petersburg, Ulyanovskaya 1, Sankt-Petersburg 198504, Russia}
\thanks{{\it E-mail:} m.ioffe@pobox.spbu.ru}
and
H. J. Korsch\address[ko]{Fachbereich Physik, Universit\"at Kaiserslautern, D-67653 
Kaiserslautern, Germany}
\thanks{{\it E-mail}: korsch@physik.uni-kl.de}}
        
\begin{document}
\maketitle
\begin{abstract}
\vspace*{4mm}
\noindent
{\bf Abstract:}\\
It is shown that the nonlinear Ermakov-Milne-Pinney equation
$\rho^{\prime\prime}+v(x)\rho=a/\rho^3$ 
obeys the property of 
covariance under a class of transformations of its  
coefficient function. This property is derived by using supersymmetric, or Darboux, transformations. The
general solution of the transformed equation is expressed in terms of the solution of the original one. Both iterations of these 
transformations 
and irreducible transformations of second order in derivatives are considered to obtain the chain of mutually related 
Ermakov-Milne-Pinney equations. 
The behaviour of the Lewis invariant and the quantum number function for bound states is investigated.
This construction is illustrated by the simple example of an infinite square 
well.\\
\end{abstract}


\vspace*{4mm}
\setcounter{footnote}{0} 
\renewcommand{\thefootnote}{\arabic{footnote}} 

\hspace*{2ex}
{\bf\large 1.} The interrelation between a special nonlinear differential equation,
known as the Ermakov, Milne or Pinney \cite{Pinn50} equation (EMP equation), and a corresponding linear equation
has been investigated in great detail by many authors
(see \cite{81mil}, the recent review in \cite{Espi00} and references therein). 

In classical dynamics these equations appear most often in context with
time-dependent oscillators in the form
\be
\ddot \rho + \omega^2(t) \rho =a\rho^{-3}
\label{1}
\ee
and
\be
\ddot q +\omega^2(t)q=0\,,
\label{haosc}
\ee
where the so-called Lewis invariant \cite{lewis}
\be
I=\frac{1}{2} \Big[ \,a \frac{q^2}{\rho^2}+(\rho \dot q-\dot \rho q)^2 \Big]
\ee
plays a  role as a constant of motion for (\ref{haosc}). For a recent application to pulse induced
transitions see \cite{sweden}. 

In Quantum Mechanics, the general solution of the Schr\"odinger equation
\be
\psi'' +k^2(x)\psi =0 \ ,\quad k^2(x)=\frac{\hbar^2}{2m}\big(E-V(x)\big)
\ee
can be expressed in amplitude-phase form as
\be
\psi(x)=\alpha \,\rho(x)\sin \Big( \int_{x_0}^x\rho^{-2}(x')\,{\rm d}x' +
\beta \Big),
\label{solschr}
\ee
where $\alpha$, $\beta$ are constants and $\rho (x)$ is an arbitrary
solution of 
\be
\rho'' + k^2(x) \,\rho =a\rho^{-3}\,.
\ee
Eq.~(\ref{solschr}) directly implies the condition
\be
N(E) = \frac{1}{\pi}\int_{-\infty}^{+\infty}\rho^{-2}(x,E)\,{\rm d}x
\stackrel{!}{=}n+1\ , \quad n=0,1,2,\ldots 
\label{Nquant}
\ee
for the bound state energies $E_n$.

In reverse, the Ermakov method can be considered as a method to solve the
EMP equation. Here we will discuss some  aspects of the
Darboux \cite{darboux}, or supersymmetric \cite{review}, covariance of this nonlinear differential equation.
The role of Darboux transformations in the study of soliton solutions of nonlinear evolutionary equations, like KdV and KP,
is well known (see, for example, \cite{salle} and references therein). Nevertheless, up to our knowledge, this approach was 
not used for an investigation of EMP equation.

\hspace*{2ex}

{\bf\large 2.} We start from the standard one-dimensional Schr\"odinger equation $(\hbar =1, m=1/2)$:
\be
-\psi^{\prime\prime} (x,E) + \biggl( W^2(x) - W^{\prime}(x) \biggr)\psi (x,E) = E\psi (x,E),
\label{schr}
\ee 
where for convenience the potential $V(x)$ is written in "supersymmetric form" \cite{review} 
in terms of the real superpotential $W(x) :$
\be
V(x) = W^2(x)-W^{\prime}(x).
\label{ricatti}
\ee 
The Hamiltonian $H = -\partial^2 + V(x)$ is intertwined with another (superpartner) Hamiltonian
\ba
\tilde H = -\partial^2 + \tilde V(x); \label{tildeH}\\
\tilde V(x) = W^2(x)+W^{\prime}(x) \label{tildeV}
\ea
by the component of supercharge
\ba
q^+ = -\partial + W(x);
\label{q+}\\
H q^+ = q^+ \tilde H.
\label{intertw}
\ea
This intertwining relation, together with the hermitian conjugated one, 
\ba
q^- H = \tilde H q^-   \label{intertww}\\
q^- \equiv (q^+)^{\dagger}=\partial + W(x), \label{q-}
\ea
seem to be the most important
elements of standard supersymmetrical Quantum Mechanics (SUSY QM) \cite{review} and its generalizations 
\cite{ais,acdi,ain,abi}. In 
particular, just these relations
(\ref{intertw}), (\ref{intertww}) lead to the isospectrality (possibly, up to zero modes of operators 
$q^{\pm}$) of
Hamiltonians $H, \tilde H$ and to the connection between their normalized eigenfunctions, if $E$ belongs to a discrete 
part of the spectrum:
\ba
\tilde\psi (x,E)=\frac{1}{\sqrt E}q^-\psi (x,E) \label{tildepsi}\\
\psi (x,E)=\frac{1}{\sqrt E}q^+\tilde\psi (x,E). \label{psi}
\ea 
Thus, if the model with Hamiltonian $H$ is exactly solvable (or 
quasi-exactly-solvable \cite{turbiner}), the model with
Hamiltonian $\tilde H$ is also exactly solvable (or quasi-exactly-solvable). By iteration of this procedure,
from $\tilde{V} = W^2+W' \equiv \tilde{W}^{2} - \tilde{W}' + c;\, c=const$ to the next potential $\tilde{\tilde{V}} = 
\tilde{W}^{2} + \tilde{W}' + c, $
one can obtain a chain of isospectral (up to zero modes of $q^{\pm}$) models with known eigenvalues and 
eigenfunctions\footnote{In notations of \cite{acdi} this is a class of so called {\it reducible} second order supertransformations 
with a shift $c.$}. In terminology of Mathematical Physics we described now schematically the well known 
Darboux-Crum \cite{crum} 
transformations \cite{salle} for Sturm-Liouville operator.

Let us denote two pairs of linearly independent solutions (may be both in 
the pair are not normalizable) of the stationary 
Schr\"odinger equations with Hamiltonians 
$H$ and $\tilde H$ by $\psi_1(x,E),\,\,
\psi_2(x,E)$ and $\tilde\psi_1(x,E), \,\, \tilde\psi_2(x,E),$ 
correspondingly. By using 
(\ref{tildepsi}), (\ref{psi}), one can check that their Wronskians coincide:
\be
\Lambda \equiv \psi_1^{\prime}\psi_2 - \psi_1\psi_2^{\prime} = \tilde\psi_1^{\prime}\tilde\psi_2 - 
\tilde\psi_1\tilde\psi_2^{\prime}\equiv \tilde\Lambda = const. 
\label{wronskian}
\ee 

As was found \cite{eliezer}, an arbitrary pair of 
linearly independent solutions $\psi_1,\,\,
\psi_2$ of linear equation (\ref{schr}) leads to a general solution 
\be
\rho(x,E)\equiv \biggl[ A \psi_1^2(x,E) + B \psi^2(x,E) + 
2C \psi_1(x,E)\psi_2(x,E) \biggr]^{1/2}
\label{rho}
\ee
of the nonlinear EMP equation (for simplicity we normalized $\rho$ in Eq.(\ref{1}) as $a=1$):
\be
-\rho^{\prime\prime}(x,E) + \biggl( W^2(x)-W^{\prime}(x,E)-E 
\biggr)\rho(x,E) = \frac{-1}{\rho^3(x,E)},
\label{milne}
\ee  
where $A,B,C$ above are arbitrary constants, which  
have to satisfy the relation
\be
AB-C^2 = \frac{1}{\Lambda^{2}}.
\label{lambda}
\ee

By a similar procedure with an arbitrary pair\footnote{We choose here the same constants $A,B,C$ and
take into account the equality (\ref{wronskian}) between Wronskians.} 
$\tilde\psi_1(x,E),\,\,\tilde\psi_2(x,E)$ of solutions of the partner 
Schr\"odinger equation with Hamiltonian (\ref{tildeH}), (\ref{tildeV}), one can construct a general solution
\be
\tilde\rho(x,E)\equiv \biggl[ A \tilde\psi_1^2(x,E) + B \tilde\psi^2(x,E) +
2C \tilde\psi_1(x,E)\psi_2(x,E) \biggr]^{1/2}
\label{tilderho}
\ee
of the partner EMP equation (with the partner coefficient 
function $(\tilde V(x) - E)$):
\be
-\tilde\rho^{\prime\prime}(x,E) + \biggl( W^2(x) + W^{\prime}(x,E)-E
\biggr)\tilde\rho(x,E) = \frac{-1}{\tilde\rho^3(x,E)}.
\label{tildemilne}
\ee

After substitution of (\ref{tildepsi}) in (\ref{tilderho}), one obtains a general solution 
$\tilde\rho(x,E)$ of (\ref{tildemilne}), which is expressed now {\it not} in terms of $\tilde\psi_1 ,\, 
\tilde\psi_2$, but in terms of linearly 
independent solutions $\psi_1,\, \psi_2$ of the {\it partner} Schr\"odinger equation (\ref{schr}). 
In such a way we found an indirect connection between solutions $\tilde\rho(x,E)$ and $\rho(x,E)$
of two EMP equations via direct connection of solutions of "their" Schr\"odinger equations. 
But this prognosticated construction would become much more elegant, if the general solution 
$\tilde\rho(x,E)$ of (\ref{tildemilne}) could be expressed directly in terms of solution $\rho(x,E)$
of (\ref{milne}). Indeed, this is the case: straightforward calculations lead to the following 
nonlinear expression:
\ba
E\tilde\rho^2(x,E) &=& \biggl(\rho^{\prime}(x,E)\biggr)^2 + 
W(x)\biggl(\rho^{2}(x,E)\biggr)^{\prime} 
+ W^2(x)\biggl(\rho (x,E)\biggr)^2 +
\frac{1}{\rho^2(x,E)} = \nonumber\\ 
&=& \biggl[\biggl(\partial + W(x)\biggr)\rho (x,E)\biggr]^2 + 
\frac{1}{\rho^2(x,E)}.  
\label{rhotilderho}
\ea 

So, the nonlinear EMP equation possesses the following property, similar to 
the supersymmetry of Schr\"odinger equation (or to Darboux covariance of
Sturm-Liouville operator): if $\rho (x,E)$ satisfies equation 
(\ref{milne}), its {\it nonlinear} transformation
(\ref{rhotilderho}) satisfies the same EMP equation, 
however with a coefficient function transformed accordingly to (\ref{tildemilne}).

By iterations one obtains again a chain of such EMP equations with related coefficient functions ("potentials"), 
whose 
general solutions can be 
expressed consecutively in 
terms of the solution $\rho (x,E)$ of the first original equation 
(\ref{milne}). This is a nonlinear analogue of 
Darboux-Crum reducible transformations for the Sturm-Liouville equation.
In particular, for two consecutive transformations:
\be
V=W^2-W^{\prime} \rightarrow \tilde V=W^2+W^{\prime} \equiv \tilde W^2 - \tilde W^{\prime}+c \rightarrow
\tilde{\tilde{V}} = \tilde W^2 + \tilde W^{\prime}+c;\quad \rho \rightarrow \tilde\rho \rightarrow \tilde{\tilde{\rho}} ,
\label{chain}
\ee
the solution $\tilde{\tilde{\rho}}$ of EMP equation (\ref{milne}) with coefficient function $\tilde{\tilde{V}}$ can be 
expressed in terms of original solution $\rho :$
\ba
E^2\tilde{\tilde{\rho}}^2 &=& \biggl[W\biggl(W+\tilde W\biggr)-E\biggr]^2 \rho^2 + \biggl(W+\tilde W\biggr)^2 (\rho^{\prime})^2 +
\biggl(W+\tilde W\biggr)\biggl[W\biggl(W+\tilde W\biggr)-E\biggr] (\rho^2)^{\prime} \nonumber\\ &+& \frac{(W+\tilde W)^2}{\rho^2},
\label{crum}
\ea
where $W(x)$ and $\tilde W(x)$ are connected by equality (\ref{chain}) and can be expressed \cite{acdi} in terms of a single
function $f(x) :$
\be
W = f - \frac{2f^{\prime}-c}{4f};\quad \tilde W = f + \frac{2f^{\prime}-c}{4f}.
\label{acdi}
\ee

The transformation (\ref{rhotilderho}) can be inverted: 
\be
E\rho^{2}(x,E) = \biggl[\biggl(\partial - W(x)\biggr)\tilde\rho (x,E)\biggr]^2 + \frac{1}{\tilde\rho^{2}(x,E)}.
\label{rhorhotilde}
\ee
Though the difference in signs in front of $W(x)$ in (\ref{rhotilderho}) and (\ref{rhorhotilde}) seems to be obvious
due to the relation between (\ref{ricatti}) and (\ref{tildeV}), the direct check of (\ref{rhorhotilde}) is not so trivial:
it is useful to explore the intermediate auxiliary equality for $\tilde\rho$ and $\rho :$  
\be
\tilde\rho\tilde\rho^{\prime} - W \tilde\rho^2 = - \rho\rho^{\prime} - W \rho^2 .
\label{equality}
\ee

{\bf\large 3.} The various generalizations of standard SUSY QM transformations were investigated in the literature. In 
particular, 
SUSY transformations of higher order in derivatives were constructed in \cite{ais,acdi,ain} (see also
the recent papers \cite{aoyama}). It was shown \cite{ais}, \cite{acdi} for the one-dimensional Schr\"odinger equation, 
that only 
two "elementary" types of transformations exist: 
the first order ones, generated by $q^{\pm}$ ( see (\ref{q+}), (\ref{q-}) ),
and the irreducible second order transformations, 
which can not be expressed as a product of two standard first order transformations.
All supercharges (and supertransformations) of higher order can be represented as a number of first- and second-order 
steps 
\cite{acdi}. 
In the context of the present paper it seems to be interesting, in addition to the construction which lead to 
(\ref{rhotilderho}) and (\ref{rhorhotilde}),  to take into account 
these second order supertransformations of Schr\"odinger
operator and their possible analogues for EMP equation.

The most general form of solution of intertwining relations (\ref{intertw}), but with second order supercharges $q^{\pm},$ was found in
\cite{acdi}:
\ba
q^+ &=& \partial^2 - 2f(x)\partial + b(x); \label{qq+}\\
q^- &=& (q^+)^{\dagger} = \partial^2 + 2f(x)\partial + 2f^{\prime}(x) + b(x); \label{qq-}\\
b(x) &=& f^{2}(x) - f^{\prime}(x) - \frac{f^{\prime\prime}(x)}{2f(x)} + \biggl( \frac{f^{\prime}(x)}{2f(x)} \biggr)^2 + \frac{d}{4f^2(x)}; 
\label{d}\\
V(x) &=& -2f^{\prime}(x) + f^2(x) + \frac{f^{\prime\prime}(x)}{2f(x)} - \biggl( \frac{f^{\prime}(x)}{2f(x)} \biggr)^2 - 
\frac{d}{4f^2(x)}; \label{VV}\\
\tilde V(x) &=& +2f^{\prime}(x) + f^2(x) + \frac{f^{\prime\prime}(x)}{2f(x)} - \biggl(\frac{f^{\prime}(x)}{2f(x)}\biggr)^2 - 
\frac{d}{4f^2(x)}\label{tildeVV},
\ea 
where $f(x)$ is 
an arbitrary real function. For reducible transformations $d=-c^2/4 ,$ it was introduced already in Eq.(\ref{acdi}), and 
an irreducible situation corresponds to the case when the {\it constant $d$ is positive}.

Substitution of\footnote{The normalization factor is obtained from the corresponding relation of the algebra of Second 
Order SUSY Quantum Mechanics \cite{ais,acdi}: $q^+q^-=H^2 + d; \,\, d>0 .$} 
\be
\tilde\psi_i(x,E) = \frac{1}{\sqrt{E^2 + d}} q^-\psi_i(x,E); \quad i=1,2 \label{tildepsipsi}
\ee
with the second order $q^-$ from (\ref{qq-}), (\ref{d}), into Eq.(\ref{tilderho}) leads to new expression for 
$\tilde\rho(x,E)$
in terms of original solutions $\psi_1(x,E),\,\psi_2(x,E)$. This expression, after replacing $\psi^{\prime\prime}$s via 
Schr\"odinger equation 
(\ref{schr}), again can be written in terms of function $\rho(x,E)$ and its first derivative only:
\ba
(E^2 + d)\tilde\rho^2(x,E) &=& 4f^{2}(x) \biggl(\rho^{\prime}(x,E)\biggr)^2 +\biggl(2f^2(x)-f^{\prime}(x)-E\biggr)^2\rho^{2}(x,E)+ 
\nonumber\\&+&
2f(x)\biggl(2f^2(x)-f^{\prime}(x)-E\biggr) \biggl(\rho^2(x,E)\biggr)^{\prime} +
\frac{4 f^2(x)}{\rho^2(x,E)}.
\label{rrhotilderho}
\ea

{\bf\large 4.} One of the reasons of the rather long peculiar interest to Ermakov systems and to related EMP equations is 
connected with existence
of the $x-$independent quantity (or $t-$independent in the context of classical harmonic oscillator instead of 
Schr\"odinger equation 
(\ref{schr})). This invariant, so called Ermakov-Lewis invariant, is expressed in terms of an arbitrary solution $\Psi (x,E)$ of 
(\ref{schr}) and an arbitrary solution $\rho (x,E)$ of the related "auxiliary" equation (\ref{milne}):
\be
I = \frac{1}{2} \biggr[\biggl(\frac{\psi (x,E)}{\rho (x,E)}\biggr)^2 + \biggl(\rho (x,E)\psi^{\prime}(x,E) - \rho^{\prime}(x,E)\psi 
(x,E)\biggr)^2\biggr].
\label{inv}
\ee

Calculation of (\ref{inv}) with $\tilde\rho$ and $\tilde\Psi$ instead of $\rho$ and $\Psi$ shows
that the value of  Ermakov-Lewis invariant
is not changed under this supersymmetric transformation and its iterations, i.e. $I = \tilde I .$

The quantum number function (\ref{Nquant}) for bound states $E_n$ transforms to an analogous function $\tilde N (\tilde E_n)$
for the system (\ref{tildeH}). $\tilde N (\tilde E_n)$ either coincides with $N(E_n)$ or differs by $\pm 1.$ 
This difference depends on the interrelation of spectra $H$ and $\tilde H ,$
i.e. \cite{abi} on the asymptotic properties of superpotential $W(x).$ For the case of second order transformations  
(\ref{qq-}) this difference can be equal to $\pm 2$ (see \cite{acdi}).

\hspace*{2ex}
{\bf\large 5.} As an illustration, we will discuss a simple example in some detail,
namely the infinite square well potential
\begin{eqnarray}
V(x)=\left\{\begin{array}{rl}-1 & \quad |x|\le \pi/2\\[2mm]
\infty & \quad |x|> \pi/2 \end{array}\right. \,,
\end{eqnarray}
which can be expressed (\ref{ricatti}) in terms of the superpotential
\begin{eqnarray}
W(x)=\left\{\begin{array}{rl}\tan x & \quad |x|\le \pi/2\\[2mm]
\infty & \quad |x|> \pi/2 \end{array}\right. \,.
\end{eqnarray}
In the following we will confine ourselves to the well region
$|x|\le \pi/2$ unless otherwise stated. 
The supersymmetric partner is:
\be
\tilde V(x)=W^2(x)+W^\prime (x)=-1+2\sec^2x
\label{Vtilde}
\ee
for the supersymmetric partner potential. The eigenvalues
for $V(x)$ are
\be
E_n=n(n+2) \ , \quad n=0,1,2,\ldots ,
\label{En}
\ee
where by construction the ground state eigenvalue $E_0$ is zero.
Because of supersymmetry, the eigenvalues coincide with those of $\tilde V(x)$:
\be
 E_{n+1} = \tilde E_n\ , \quad n=0,1,2,\ldots \,.
\label{EE}
\ee
According to Eq.~(\ref{rho}) we can now write the general solution
of the EMP-equation (\ref{milne})
\be
\rho^{\prime\prime}(x) + k^2\rho(x) = \rho^{-3}(x),\quad  k^2=E+1 
\ee
in terms of the solutions $\psi_1(x)=\sin kx$ and  
$\psi_2(x)=\cos kx$
of the Schr\"odinger equation:
\be
\rho^2(x)= A \sin^2 kx + B \cos^2 kx + 2C \sin kx \cos kx
\, , \quad AB-C^2=1/k^2\,.
\ee
Note that $\rho(x)$ is infinite outside the potential well.
Assuming the special case $C=0$, $A=B=1/k$ for simplicity, we have
the solution
\be
\rho^2(x)=1/k = const.
\label{rho2-ex}
\ee
As a check of the Milne quantization condition (\ref{Nquant})
one obtains
\be
N(E) = \frac{1}{\pi}\int_{-\infty}^{+\infty}\rho^{-2}\,{\rm d}x
= \frac{1}{\pi}\int_{-\pi/1}^{+\pi/2}k\,{\rm d}x = k
\stackrel{!}{=}n+1\ , \quad n=0,1,2,\ldots
\label{Nex}
\ee
which reproduces, of course, Eq.(\ref{En}) 

By means of the supersymmetric covariance we can now immediately
write down the solution (\ref{rhotilderho}) of the EMP-equation (\ref{tildemilne}) for the partner potential
(\ref{Vtilde}):
\be
\tilde \rho^2(x)=\big( \rho^\prime +\rho  \tan x\big)^2+\rho^{-2}=
k^{-1}\tan^2 x +k=\gamma \big(k^2+\tan^2x\big)
\label{rhotilde-ex}
\ee
with $\gamma = 1/kE$.

By some elementary algebra, one can directly check that $\tilde\rho$ is indeed a solution of the EMP equation (\ref{tildemilne}).

From the solution (\ref{rhotilde-ex}) we can also compute the
quantum number function
\be
\tilde N(E) = \frac{1}{\pi}\int_{-\infty}^{+\infty}
\frac{{\rm d}x}{\tilde\rho^2}
=\frac{1}{\pi \gamma}\frac{\pi}{k(1+k)}
=k-1
\ee
and with $N(E)=k$ from (\ref{Nex}) we arrive at
the relation $N(E)=\tilde N(E)+1$ as was predicted, taking into account Eq.(\ref{EE}).

Finally, we can also verify that the values of the invariants
$I$ and $\tilde I$ agree. If $\Psi (x)$ is any solution of
the Schr\"odinger equation for the square well potential
$V(x)$ and $\rho^2(x)=1/k$ is taken from (\ref{rho2-ex}),
we have
\be
2I=\Big(\frac{\Psi}{\rho}\Big)^2+(\rho \Psi^\prime - \rho^\prime \Psi)^2
=k\Psi_0^2+k^{-1}\Psi^{\prime \,2}.
\ee
The partner wave function is $\sqrt{E}\,\tilde \Psi =\Psi^\prime +\tan 
x\,\Psi$
with $\sqrt{E}\,\tilde \Psi^\prime= (\tan^2x-E)\,\Psi +\tan 
x\,\Psi^\prime$.
We evaluate the (constant) value of the invariant $\tilde I$ at
$x=0$, which gives with $\Psi(0)=\Psi_0$ etc. and
$\tilde \rho_0^2=\gamma k^2$, $\tilde \rho^\prime_0 =0$\,:
\begin{eqnarray}
2\tilde I&=& \Big(\frac{\tilde \Psi_0}{\tilde \rho_0}\Big)^2+
\big(\tilde \rho_0 \tilde \Psi^\prime_0-
\tilde \rho^\prime_0 \tilde \Psi_0\big)^2\nonumber\\[2mm]
&=&\Big(\frac{\Psi_0^\prime}{\sqrt{E}}\Big)^2\,\frac{1}{\gamma k^2} +
\gamma k^2\big( -\sqrt{E}\,\Psi_0\big)^2
=k^{-1}\Psi^{\prime \,2}_0+ k\Psi_0^2 =2I\,.
\end{eqnarray}

\section*{\large\bf\quad Acknowledgments.}
\hspace*{3ex}
One of the authors (M.I.) is grateful to DAAD for support of this work and 
to the University of Kaiserslautern, where this work was done, 
for warm hospitality.
The work of M.I. was also partially supported by RFBR grant N 02-01-00499.

\end{document}